# Modeling and Validation of Electrical Load Profiling in Residential Buildings in Singapore

Luo Chuan and Abhisek Ukil, *Senior Member, IEEE*

*Abstract*—The demand of electricity keeps increasing in this modern society and the behavior of customers vary greatly from time to time, city to city, type to type, etc. Generally, buildings are classified into residential, commercial and industrial. This study is aimed to distinguish the types of residential buildings in Singapore and establish a mathematical model to represent and model the load profile of each type. Modeling household energy consumption is the first step in exploring the possible demand response and load reduction opportunities under the smart grid initiative. Residential electricity load profiling includes the details on the electrical appliances, its energy requirement, and consumption pattern. The model is generated with a bottom-up load model. Simulation is performed for daily load profiles of 1 or 2 rooms, 3 rooms, 4 rooms and 5 rooms public housing. The simulated load profile is successfully validated against the measured electricity consumption data, using a web-based Customer Energy Portal (CEP) at the campus housings of Nanyang Technological University, Singapore.

## I. Introduction

IN the absence of natural resources for the generation of electricity, and with the increasing population and energy demand, energy is one of the critical factors for the development of Singapore's economy in the immediate and long-term future. With the growing awareness of environmental issue, many studies are focusing on smart technologies to improve the energy utilization efficiency, e.g., in the smart grid framework. It is known that smart grid is based on the behavior of customers to control the supply and the load in the power system. Therefore, a good understanding and estimation of customer load profiling is the fundamental for the smart grid enabling technologies, in the low voltage (LV, <1 kV) distribution side. On the other hand, load data is crucial for planning electricity distribution networks and optimal production capacity. Accurate knowledge of the household consumer loads is important when small scale distributed energy technologies are optimally sized into the local distribution network, with demand side management (DSM) measures.

According to the energy statistics [1] carried out by the Energy Market Authority (EMA), Singapore in the year 2013, the total electricity consumed by buildings (residential, commercial and industrial) occupied 93.3% of the overall energy consumption in Singapore. Residential buildings refer to the public and the private housings. Commercial or office buildings refer to those used for rendering services like agency, commission, banking, administrative, legal, architectural, and other professional services like shopping malls, etc. Industrial buildings refer to factories, production units, etc., mostly at medium voltage (MV, <35 kV) as well as low voltage (LV, <1 kV) levels.

As shown in Fig. 1 [1], contestable customers refer to those electricity customers whose energy consumption is high enough to be allowed to purchase electricity either from third-party retailers or the wholesale market, whereas non-contestable customers refer to those who can only buy electricity from the wholesale market.

It is shown that households' energy consumption account for 47.3% of the non-contestable load which is equivalent to 15.6% of the total load. Commercial buildings used 37.9% of the entire consumed energy in 2013 and 39.8% of the total electricity energy was consumed by industrial related activities [1].

This study reports the detailed load profiling of the residential buildings in Singapore, along with the mathematical models, with the following objectives:
1) to classify the types of buildings;
2) to study the average load of each type of residential buildings in Singapore;
3) to study the bottom-up model and fit it into residential building load profile;
4) to construct the residential load from elementary load components;
5) to validate the modeled load profile using measured data.

The remainder of the paper is organized as follows. Section II presents the literature review on energy consumption modeling in residential buildings, bottom-up modeling. Section III presents the load profiling, covering the factors influencing energy utilization, mathematical models. Detailed simulation results for the load profiles in public housing with 1 or 2, 3, 4 and 5 rooms are described in Section IV. Section V shows the validation of the load profile models against the measured data. Discussions on the proposed method and the results are described in Section VI, followed by conclusions in Section VII.

The authors are with the School of Electrical & Electronic Engineering, Nanyang Technological University, Singapore (e-mail: CLUO002@e.ntu.edu.sg; aukil@ntu.edu.sg).

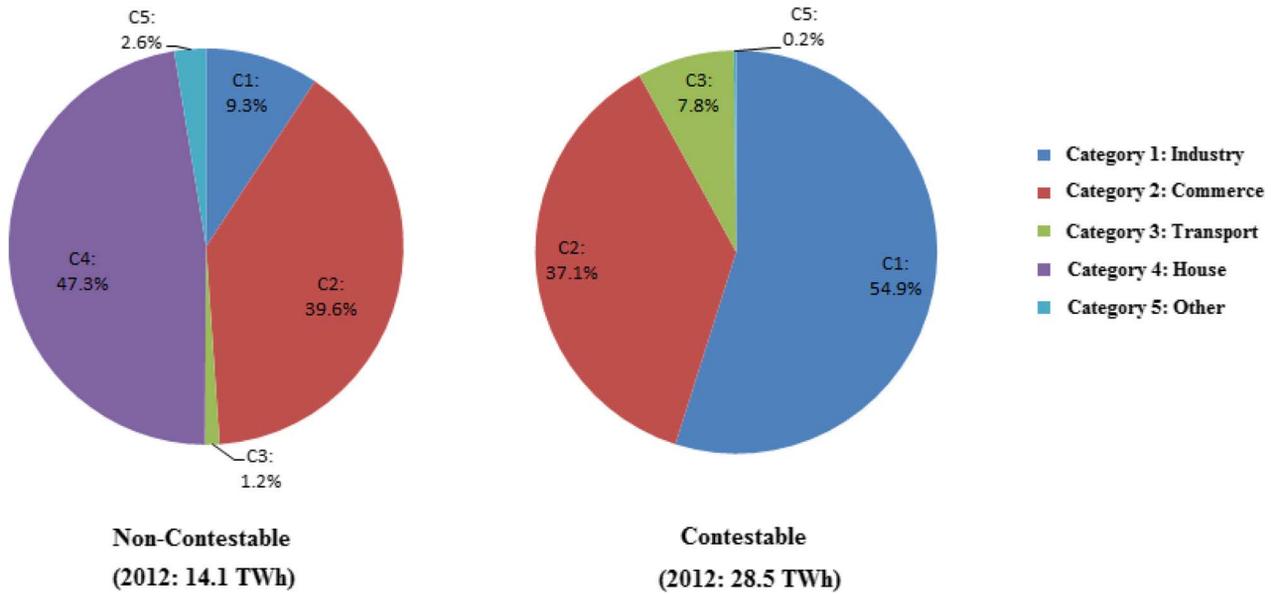

Fig. 1. Electricity sales by contestability and sectors in Singapore [1].

## II. Literature Review

### A. Residential Building Electricity Consumption in Singapore

From the report by EMA, Singapore, energy consumption in the residential buildings experienced an increase at 4.9% in the one year period and reached 15.6% of the entire electricity load [1].

For household electricity users, the average monthly electricity consumption can be categorized according to the type of dwelling. As shown in Fig. 2—plot (i), a typical public housing 4-room unit consumed about 379.6 kWh of electricity a month in 2012. Being the most common type of public housing, as shown in Fig. 2—plot (ii), 4-room unit can be taken as benchmark [2]. Using this as a benchmark, 1–2 room and 3-room units on average consumed 40.5% (or 153.8 kWh) and 73.3% (or 278.2 kWh) of electricity relative to a 4-room unit respectively [1]. Average electricity consumption in a public housing 5-room and executive flat (465.0 kWh) was 22.5% higher compared to a typical 4-room flat [3].

### B. Effect of Ambient Temperature

Another factor that affects the residential electricity consumption is the ambient temperature. Fig. 3 shows the relation between the temperature and the residential electricity consumption in the year 2011. From Fig. 3, it can be observed that the peak and the bottom parts of the electricity consumption curve follow the same trend of the temperature change. The peak electricity load occurred in June when the average temperature is also among the highest in the year. The lowest consumption was observed in February and March which coincidently happened to be the coolest period in the year.

However, the fluctuation of the energy consumption between the peak and the bottom part is only around 50 kWh which is not considered as a very significant change. This is because the characteristic of Singapore's climate is tropical marine type, with a relatively stable and constant temperature profile throughout the year. Therefore, in this study focused on Singapore, the effect of temperature to the residential load is excluded.

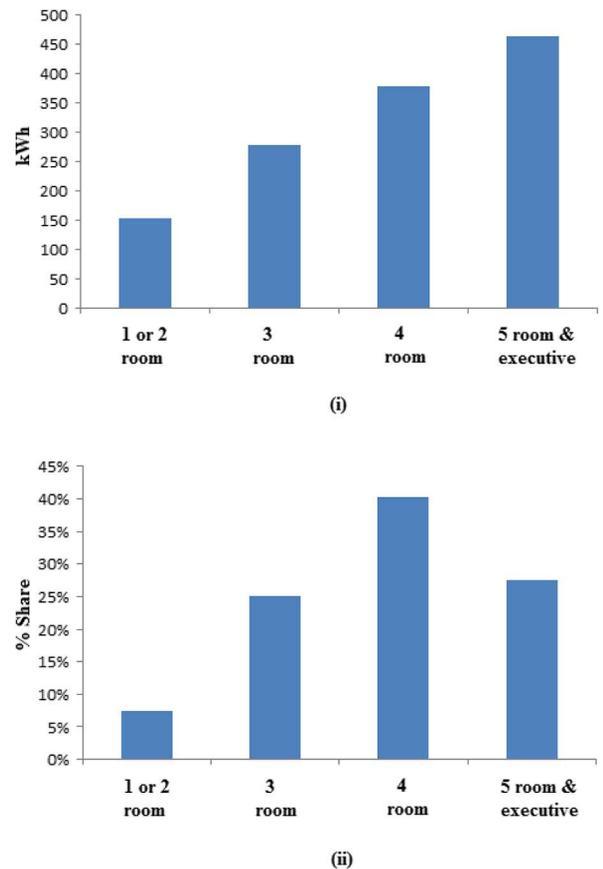

Fig. 2. Plot (i): Average monthly electricity consumption by dwelling type [1]. Plot (ii): Percentage share of dwelling types in housing market according to number of rooms.

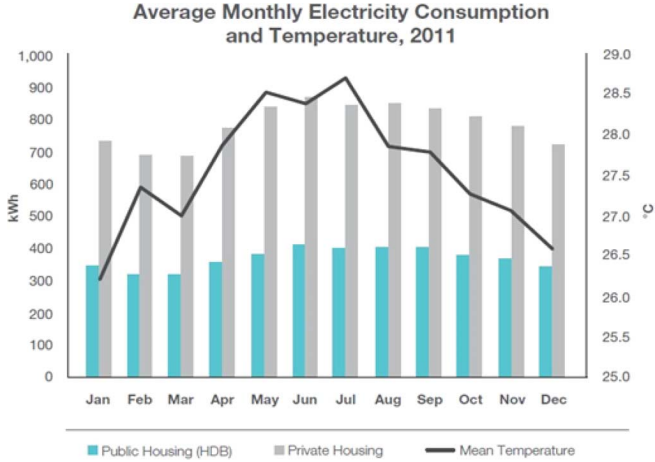

Fig. 3. Average monthly electricity consumption and ambient temperature.

### C. Bottom-Up Approach for Modeling

A number of researches [4]–[15] were done on the topic of electricity demand modeling at the utility level. Gross and Galiana reported review of short-term load forecasting methods in [4]. A number of reviews on utilization of neural networks for short-term load forecasting are reported in [5] and [6]. Forecasting methods using fuzzy logic, neural networks, genetic algorithm and expert systems are described in [7] and [8].

Forecasting for different countries is reported as well. For example, Lee *et al.* reported short-term load forecasting in South Korea [9], He *et al.* reported case study in China [10], Ning *et al.* reported case study in the USA [11]. Above-mentioned studies and methods are commonly employed when there is lack of, or no knowledge about the appliance stocks and other grass-root level consumer details.

In the works [12]–[15], an end-user model using the bottom-up approach was introduced and proved to be a reliable method to simulate the household electricity consumption. Basically the idea of bottom-up approach is to build up the total load from the elementary load components which can be a single household or even a single piece of appliance. The advantage of bottom-up approach is that it can analyze every individual appliance's effect on the total load which could be quite helpful in the future study of smart grid. The logic of the bottom-up approach is shown in Fig. 4, and followed in this study.

As shown in Fig. 4, the point of entry to the procedure is marked with a large triangle pointing towards the next block while the exit point is marked with a large triangle pointing out. The parts including computational loops have two exit arrows, one with solid line and one with dashed line. The solid line points to the repeating loop itself, while the dashed line points to the following step, as the loop exits.

When the type of the household is given, the appliance load curve loop is activated. The set of appliance used are defined statistically. The hourly power consumed by the individual appliance is estimated and fed back into the overall household load curve to generate a total household load by accumulating all individual appliance load curve.

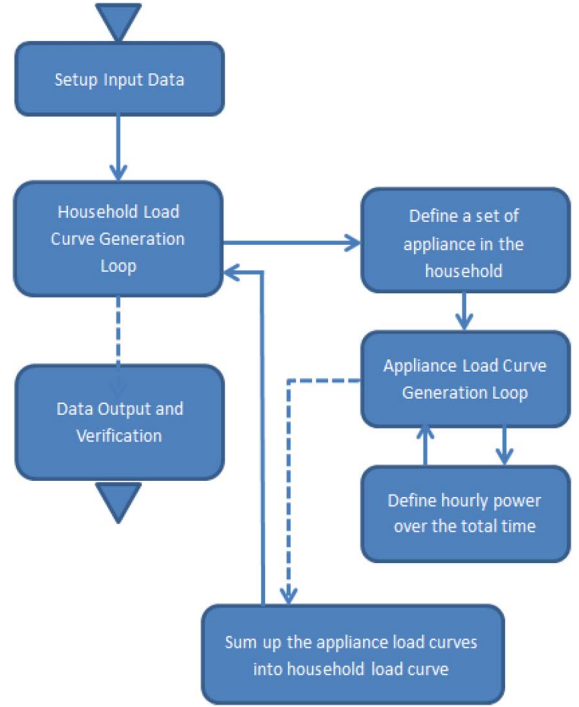

Fig. 4. Diagram for bottom-up load generation procedure.

A typical limitation for detailed bottom-up methods is an extensive need of data about the consumers or their appliances and the households. Usually, some parts of the data are not easily available. In the Capasso model [12], detailed data is needed about consumer behavior. In addition to the consumer behavior, the Norwegian ERAD model [15] also requires very detailed information about the design of the house in which the household is located. In our work, the need for detailed data is bypassed by using a representative data sample and statistical averages.

## III. RESIDENTIAL BUILDING LOAD PROFILE

The average daily, monthly and yearly Singapore residential electricity usage, accessible from EMA official website [16], is normally aggregated consumption of multiple households, without knowledge about the events in individual households. There is no available data that could provide information about how the domestic electricity load is composed of different electricity appliances in an individual household and the utilization pattern. In this section, the public residential households' load is simulated according to the house type. The analysis method employed in this research is the bottom-up approach which builds up the total load for each type of house, considering every piece of home appliance, in a statistical average manner.

### A. Factors Influencing Energy Utilization

A number of researchers have studied the residential household electricity load characteristics and discovered some periodical patterns on a seasonal, daily and hourly scale [17]–[20]. In 2002 and 2003, two sets of domestic consumption data were collected in Finland with a sample size of around 1800 households in total. Further studies on the data show that the daily electricity consumption on a yearly level is often dependent on external variables such as the mean outside temperature and daily daylight hours that typically follow similar patterns over successive years, known as seasonal effect.

However, as the daily temperature pattern and daylight hours change slightly in Singapore (see Section II-B), the daily electricity consumption on a yearly level can be considered as consistent with insignificant change. Therefore, the seasonal effect is excluded in this study.

Previous studies [17]–[20] also discovered that the hourly fluctuation of domestic loads result from the combined effect of consumer availability and activity level. Thus, the mean daytime consumption during workdays is typically lower than that in the weekends, and in the workdays evening the consumption is higher than in weekend evenings. However, due to time and resource constraint, the load discrepancy between the weekday and the weekend is ignored in this study. This will be part of future studies.

Besides the above-mentioned factors, the consumption behavior can also be affected by some uncertain factors such as abnormal weather, national or international events, and so on. For instance, a popular television show or commercial breaks can make variations observed in the national network. Such a level of details are not included into consideration in this study, which is more focused on an average and universal electricity consumption behavior in the residential households.

### B. Mathematical Model for Household Electricity Consumption

Following Fig. 4, the daily household load curve is generated through two repeating loops. The "Appliance load curve generation loop" is for the sake of generating the load curve for one day period for one particular appliance. The "Household Load Curve Generation Loop" is tasked to repeat the process until all appliances in the house are counted.

In the left part of the procedures (see Fig. 4), the model is repeated for all appliances in a household. The list of appliances varies in different households, therefore a coefficient named 'saturation level' is introduced as availability probabilities ($P_{sat}$, value between 0 and 1) of an unique set of appliance in a household. This can make the availability statistics to become coherent with the original one when a large number of households are considered. Such kind of saturation level can be found in many studies for different countries such as USA [21] and Finland [22]. In the following part, a method of deducing the appliance saturation level for Singapore's residential buildings will be introduced, and verified by the load profile.

Assuming the appliance list and corresponding saturation level is known, a mathematical model is required to build up a household's daily load profile. The idea is that when an appliance is activated, the rated load of the appliance will be added into the household's electricity consumption at the corresponding time until it completes a full cycle of a single activation. When every piece of appliance in the unit is considered, a daily load profile can be drawn by summing up the individual appliance's consumption.

An appliance can be activated at any time and multiple times in a day, depending on the usage pattern. This is the most challenging part of the simulation. In this study, a starting probability function $P_{start}$ is introduced in order to determine when an appliance will be activated. $P_{start}$ is defined for each time step, receiving a value between 0 and 1. The value of $P_{start}$ varies through a series of calculations. When the appliance is off, the turning on is checked using $P_{start}$. Activation occurs when $P_{start}$ is larger than a randomly generated number between 0 and 1 by computer. Then the consumption cycle of the appliance will be added to the household's total load curve. When the end of one consumption cycle is met, the appliance is turned off, and the checking for starting the appliance will carry on again. Starting Probability function $P_{start}$ is calculated by the following function:

$$P_{start}(A, \Delta t_{comp}, h) = P_{hour}(A, h) \times f(A, d) \times P_{step}(\Delta t_{comp}) \times P_{sat}(A). \quad (1)$$

$P_{start}$ depends on three variables, $A$ the appliance, $\Delta t_{comp}$ the computational time step in minute and $h$ is the hour of the day. $P_{sat}$ is the appliance saturation level which is introduced above. $P_{hour}$ is the hourly probability factor which models the activity levels of the appliance during a day, $f$ is the mean daily starting frequency, modeling the average time of use for an appliance and its unit is $day^{-1}$, $P_{step}$ is the step size scaling factor which scales the probabilities according to $\Delta t_{comp}$.

The hourly probability factor $P_{hour}$ is a statistical probability factor that describes the activity level of an appliance in hourly basis. Higher $P_{hour}$ value means higher chance for the appliance to be turned on, and vice versa. Compared to the actual appliance utilization statistics, $P_{hour}$ value is likely to be available more [21]–[23], which could be incorporated into this study in Singapore. The mean daily starting frequency $f$ depends on the type of household.

### C. Simulation Procedure

Equation (1) is applied for each appliance in the household. A random number between 0 and 1 is generated and compared with $P_{start}$ at the beginning of each computational time step. If $P_{start}$ wins the comparison, the appliance is turned on. When the time of the on-cycle is reached, i.e., $t = t_{start} + t_{cycle}$, the appliance is turned off and starts to check for the next turn on using the same method. If an appliance has standby electricity use, it will be added to the whole load curve continuously, for example, refrigerator and TV. $t_{cycle}$ is the average on-cycle for an appliance, which can be found from previous studies [20].

The daily average energy consumption by a household can be calculated by active and standby consumption parameters using the following formula:

$$E_{monthly} = \frac{[3600 \times W_{standby} + f \sum_{n=1}^{n_{app}} W_{nom} \times t_{cycle}] \times 30}{3.6 \times 10^6}$$
$$kWh/month. \quad (2)$$

TABLE I
RESIDENTIAL APPLIANCES: LIST, SATURATION LEVEL, NOMINAL WATTAGE, STANDBY POWER, AVERAGE DAILY STARTING FREQUENCY, AND TIME PER CYCLE

| Appliance | Appliance Saturation | Nominal Wattage (W) | Standby (W) | Mean daily starting frequency (f) | Time per cycle (min) |
|---|---|---|---|---|---|
| Microwave oven | 0.93 | 1500 | N.A. | 5 | 5 |
| Refrigerator | 1 | 110 | 8.1 | 40.5 | 12 |
| 2nd Refrigerator | 0.15 | 110 | 8.1 | 40.5 | 12 |
| Coffee Maker | 0.34 | 1000 | N.A. | 0.76 | 6 |
| Range Oven | 0.42 | 1050 | 8 | 0.46 | 12 |
| Clothes-washer | 0.97 | 1200 | N.A. | 0.36 | 54 |
| TV | 1 | 105 | 4 | 1.62 | 90 |
| 2nd TV | 0.21 | 83 | 4 | 0.28 | 60 |
| Play station | 0.26 | 96 | 2 | 0.3 | 60 |
| Computer | 0.92 | 110 | 2.5 | 2.5 | 60 |
| Air Conditioning | 0.88 | 1300 | N.A. | 1.36 | 120 |
| Hair dryer | 0.93 | 1600 | N.A. | 1.46 | 7 |
| Lighting | 1 | 120 | N.A. | 10 | 30 |

$W_{nom}$ and $W_{standby}$ are nominal and standby power in watts. $n_{app}$ is the total number of appliances. The mean daily starting frequency is denoted by $f$, and $t_{cycle}$ is the average on-cycle for an appliance.

The mathematical models are simulated using MATLAB, with the following assumptions.

1) Thirty different load profiles are not simulated; rather single day is simulated, and multiplied by 30. The mean daily starting frequency factor is assumed to incorporate the dynamics.
2) Certain appliances like the refrigerator are known to have varying power usage values throughout the day. Actions such as frequent opening and closing the doors or ice making, dispensing and crushing, etc. can effectively fluctuate the energy consumption. Other devices like air conditioning systems, etc. also get affected by users' behaviors. However, it is very difficult to model such stochastic human behavior as a general parameter. The average on-cycle for an appliance is assumed to cover up that.
3) Differentiation between weekdays and weekends is not considered.
4) The public housing size is assumed to be a good indicator of demographic and economical conditions in Singapore [2]. Extreme variations in economical conditions, influencing the appliance type and usage patterns are not considered.
5) The load models and usage patterns are assumed to be uninfluenced by the time-of-use tariff structure, as currently (relative to the time of writing of this paper) such tariff structure is not used in the residential sector in Singapore.

## IV. LOAD PROFILE SIMULATION

In this section, the detailed simulation of the load profile would be described. For brevity purpose, detailed simulation of the load profile for 1 or 2 rooms flat and 5 rooms flat would be shown, covering the lower and higher side of the residential units. Simulation of 3 and 4 rooms flats are described concisely.

### A. Residential Appliance Data

Following the study carried out in California, USA in 2009 [21], the list of appliance in Singapore households and the corresponding saturation are established, as shown in the columns 1 and 2 in Table I. Typically, the appliance saturation is calculated based on surveys in particular localities, considering the ratio of the total number of each type of appliances per households and the total number of houses. Since the appliance saturation level depends on the household type, the content in the column 2 of Table I is just a reference for the detailed investigation based on different type of units, described later.

The nominal wattage rating for the listed appliance can be found from the internet, for example reference [23] provides all the working power for above appliances. Ajay *et al.* studied the standby power consumption for computer, television and refrigerator in [24]. That information is incorporated in this study. Columns 3 and 4 in Table I respectively show the nominal wattage and the standby power of the residential appliances.

The average operating time per cycle can be deduced based on experience and it is independent of household type. However, mean daily starting frequency ($f$) is based on statistics and varies between different types of household. For example, the mean daily starting frequency of a microwave oven in a 5-room household is most likely higher than a 2-room flat. Columns 5 and 6 in Table I respectively show the typical mean daily starting frequency and the average operating time per cycle [21], [22].

### B. Load Profiling of One or Two Rooms Public Housing

The mean daily starting frequency and saturation level for one or two room houses is shown in Table II, and the starting probability $P_{start}$ is shown in Table III for 24 h.

With the above data, the simulated daily load profile for 1 or 2 rooms' flat is shown in Fig. 5, giving an average value of 148.4 kWh. From EMA report [1], the average monthly electricity consumption in 1 or 2 rooms flat is 153.8 kWh (see Section II-A), which is close to the simulated model.

TABLE II
MEAN DAILY STARTING FREQUENCY AND SATURATION
LEVEL FOR 1 OR 2 ROOMS FLAT

| Appliance | Mean daily starting frequency (f) | Appliance Saturation |
|---|---|---|
| Microwave oven | 3 | 0.93 |
| Refrigerator | 40.5 | 1 |
| Coffee Maker | 0.45 | 0.2 |
| Range Oven | 0.31 | 0.42 |
| Clothes-washer | 0.21 | 0.97 |
| TV | 1.2 | 1 |
| 2nd TV | 0.28 | 0.05 |
| Play station | 0.3 | 0.16 |
| Computer | 1.6 | 0.92 |
| Air Conditioning | 0.75 | 0.71 |
| Hair dryer | 0.9 | 0.88 |
| Other Occasional Loads | 0.41 | 1 |
| Lighting | 8 | 1 |
| Lighting | 8 | 1 |
| Lighting | 8 | 1 |

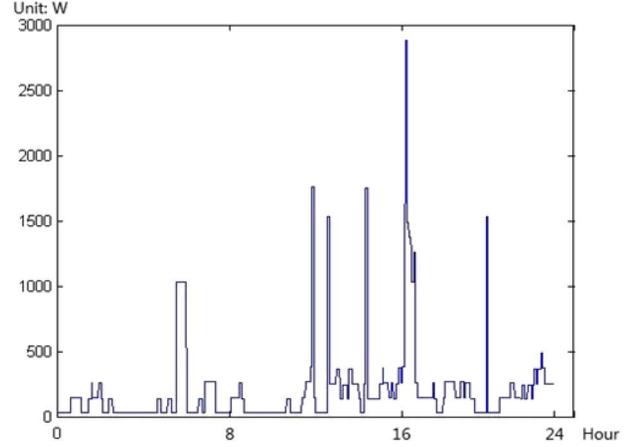

Fig. 5. Daily load profile of one or two rooms flat.

TABLE III
STARTING PROBABILITY FOR 1 OR 2 ROOMS FLAT

| Appliance | P start | | | | | | | |
|---|---|---|---|---|---|---|---|---|
| | 1 | 2 | 3 | 4 | 5 | 6 | 7 | 8 |
| Microwave oven | 0.00086 | 0.000116 | 0 | 0 | 0 | 0.000395 | 0.003999 | 0.005929 |
| Refrigerator | 0.140738 | 0.140738 | 0.140738 | 0.003475 | 0.140738 | 0.140738 | 0.140738 | 0.140738 |
| Coffee Maker | 2.78E-05 | 3.75E-06 | 0 | 0 | 0 | 1.28E-05 | 0.000129 | 0.000191 |
| Range Oven | 4.01E-05 | 5.43E-06 | 0 | 0 | 0 | 1.84E-05 | 0.000187 | 0.000277 |
| Clothes-washer | 8.49E-05 | 0 | 0 | 0 | 0 | 0 | 0 | 0.000119 |
| TV | 0.0034 | 0.00194 | 0.00087 | 0.000642 | 0.00087 | 0.00097 | 0.00097 | 0.00146 |
| 2nd TV | 3.97E-05 | 2.26E-05 | 1.02E-05 | 3.21E-05 | 1.02E-05 | 1.13E-05 | 1.13E-05 | 1.7E-05 |
| Play station | 0.000136 | 7.76E-05 | 3.48E-05 | 0.000103 | 3.48E-05 | 3.88E-05 | 3.88E-05 | 5.84E-05 |
| Computer | 0.004171 | 0.00238 | 0.001067 | 0.00059 | 0.001067 | 0.00119 | 0.00119 | 0.001791 |
| Air Conditioning | 0.001132 | 0.00059 | 0.000546 | 0.000728 | 0.00059 | 0.000768 | 0.000945 | 0.001575 |
| Hair dryer | 0.002244 | 0.00128 | 0.000574 | 0.000565 | 0.000574 | 0.00592 | 0.00724 | 0.006244 |
| Other Occasional Loads | 0.000871 | 0.000454 | 0.00042 | 0.001025 | 0.000454 | 0.000591 | 0.000728 | 0.001213 |
| Lighting | 0.017 | 0.008867 | 0.0082 | 0.00615 | 0.008867 | 0.0102 | 0.0142 | 0.027 |
| Lighting | 0.017 | 0.008867 | 0.0082 | 0.00615 | 0.008867 | 0.0102 | 0.0142 | 0.027 |
| Lighting | 0.017 | 0.008867 | 0.0082 | 0.00615 | 0.008867 | 0.0102 | 0.0142 | 0.027 |

| Appliance | P start | | | | | | | |
|---|---|---|---|---|---|---|---|---|
| | 9 | 10 | 11 | 12 | 13 | 14 | 15 | 16 |
| Microwave oven | 0.009626 | 0.013811 | 0.016205 | 0.01781 | 0.018414 | 0.016624 | 0.014857 | 0.013694 |
| Refrigerator | 0.140738 | 0.140738 | 0.140738 | 0.140738 | 0.140738 | 0.140738 | 0.140738 | 0.140738 |
| Coffee Maker | 0.000311 | 0.000446 | 0.000523 | 0.000575 | 0.000594 | 0.000536 | 0.000479 | 0.000442 |
| Range Oven | 0.000449 | 0.000644 | 0.000756 | 0.000831 | 0.000859 | 0.000776 | 0.000693 | 0.000639 |
| Clothes-washer | 0.00034 | 0.000783 | 0.001192 | 0.001227 | 0.001227 | 0.001246 | 0.001246 | 0.001246 |
| TV | 0.00243 | 0.0034 | 0.00388 | 0.00485 | 0.00485 | 0.00593 | 0.00613 | 0.0068 |
| 2nd TV | 2.84E-05 | 3.97E-05 | 4.53E-05 | 5.66E-05 | 5.66E-05 | 6.92E-05 | 7.15E-05 | 7.93E-05 |
| Play station | 9.72E-05 | 0.000136 | 0.000155 | 0.000194 | 0.000194 | 0.000237 | 0.000245 | 0.000272 |
| Computer | 0.002981 | 0.004171 | 0.004759 | 0.005949 | 0.005949 | 0.007274 | 0.007519 | 0.008341 |
| Air Conditioning | 0.001806 | 0.001771 | 0.001673 | 0.001762 | 0.001806 | 0.001984 | 0.002205 | 0.002676 |
| Hair dryer | 0.001604 | 0.002244 | 0.001241 | 0.001221 | 0.001881 | 0.001274 | 0.000746 | 0.001188 |
| Other Occasional Loads | 0.001391 | 0.001363 | 0.001288 | 0.001356 | 0.001391 | 0.001527 | 0.001698 | 0.00206 |
| Lighting | 0.020467 | 0.0266 | 0.0218 | 0.0188 | 0.017133 | 0.0218 | 0.026467 | 0.023333 |
| Lighting | 0.020467 | 0.0266 | 0.0218 | 0.0188 | 0.017133 | 0.0218 | 0.026467 | 0.023333 |
| Lighting | 0.020467 | 0.0266 | 0.0218 | 0.0188 | 0.017133 | 0.0218 | 0.026467 | 0.023333 |

| Appliance | P start | | | | | | | |
|---|---|---|---|---|---|---|---|---|
| | 17 | 18 | 19 | 20 | 21 | 22 | 23 | 24 |
| Microwave oven | 0.015764 | 0.017228 | 0.017019 | 0.01681 | 0.016112 | 0.009509 | 0.005348 | 0.002372 |
| Refrigerator | 0.140738 | 0.140738 | 0.140738 | 0.140738 | 0.140738 | 0.140738 | 0.140738 | 0.140738 |
| Coffee Maker | 0.000509 | 0.000556 | 0.000549 | 0.000542 | 0.00052 | 0.000307 | 0.000173 | 7.65E-05 |
| Range Oven | 0.000736 | 0.000804 | 0.000794 | 0.000784 | 0.000752 | 0.000444 | 0.00025 | 0.000111 |
| Clothes-washer | 0.001261 | 0.001261 | 0.001314 | 0.001314 | 0.001261 | 0.001039 | 0.000662 | 0.000153 |
| TV | 0.0068 | 0.0068 | 0.00777 | 0.00825 | 0.0068 | 0.00534 | 0.00485 | 0.00387 |
| 2nd TV | 7.93E-05 | 7.93E-05 | 9.07E-05 | 9.63E-05 | 7.93E-05 | 6.23E-05 | 5.66E-05 | 4.52E-05 |
| Play station | 0.000272 | 0.000272 | 0.000311 | 0.00033 | 0.000272 | 0.000214 | 0.000194 | 0.000155 |
| Computer | 0.008341 | 0.008341 | 0.009531 | 0.01012 | 0.008341 | 0.00655 | 0.005949 | 0.004747 |
| Air Conditioning | 0.002805 | 0.003035 | 0.003257 | 0.003355 | 0.003013 | 0.00296 | 0.002148 | 0.001429 |
| Hair dryer | 0.002508 | 0.003168 | 0.005128 | 0.005445 | 0.004488 | 0.003524 | 0.003201 | 0.002554 |
| Other Occasional Loads | 0.002159 | 0.002337 | 0.002508 | 0.002583 | 0.00232 | 0.002279 | 0.001654 | 0.0011 |
| Lighting | 0.040133 | 0.051267 | 0.0556 | 0.057067 | 0.0576 | 0.054467 | 0.043267 | 0.028333 |
| Lighting | 0.040133 | 0.051267 | 0.0556 | 0.057067 | 0.0576 | 0.054467 | 0.043267 | 0.028333 |
| Lighting | 0.040133 | 0.051267 | 0.0556 | 0.057067 | 0.0576 | 0.054467 | 0.043267 | 0.028333 |

TABLE IV
MEAN DAILY STARTING FREQUENCY AND SATURATION
LEVEL FOR 5 ROOMS FLAT

| Appliance | Mean daily starting frequency (f) | Appliance Saturation |
|---|---|---|
| Microwave oven | 7.5 | 1 |
| Refrigerator | 40.5 | 1 |
| 2nd Refrigerator | 40.5 | 0.31 |
| Coffee Maker | 1.12 | 0.37 |
| Range Oven | 0.56 | 0.49 |
| Clothes-washer | 0.75 | 1 |
| TV | 1.95 | 1 |
| 2nd TV | 0.28 | 0.39 |
| Play station | 0.3 | 0.33 |
| Computer | 3.9 | 0.92 |
| Air Conditioning | 2.13 | 0.93 |
| Hair dryer | 1.8 | 1 |
| Other Occasional Loads | 0.72 | 1 |
| Lighting | 18 | 1 |
| Lighting | 18 | 1 |
| Lighting | 18 | 1 |
| Lighting | 18 | 1 |
| Lighting | 18 | 1 |
| Lighting | 18 | 1 |
| Lighting | 18 | 1 |

### C. Load Profiling of Five Rooms Public Housing

The mean daily starting frequency and saturation level for five room houses is shown in Table IV, and the starting probability $P_{start}$ is shown in Table V for 24 h.

With the above data, the simulated daily load profile for 5 rooms flat is shown in Fig. 6, giving an average value of 447.8 kWh. From EMA report [1], the average monthly electricity consumption in 5 rooms flat is 465 kWh (see Section II-A), which is close to the simulated model.

### D. Load Profiling of Three Rooms Public Housing

For brevity purpose, detailed analysis of three rooms flats is not shown, only the daily load profile is shown in Fig. 7, giving an average value of 284.2 kWh. From EMA report [1], the average monthly electricity consumption in 3 rooms flat is 278.2 kWh (see Section II-A), which is close to the simulated model.

TABLE V
STARTING PROBABILITY FOR 5 ROOMS FLAT

| Appliance | P start | | | | | | | |
|---|---|---|---|---|---|---|---|---|
| | 1 | 2 | 3 | 4 | 5 | 6 | 7 | 8 |
| Microwave oven | 0.002313 | 0.000313 | 0 | 0 | 0 | 0.001063 | 0.01075 | 0.015938 |
| Refrigerator | 0.140738 | 0.140738 | 0.140738 | 0.003475 | 0.140738 | 0.140738 | 0.140738 | 0.140738 |
| 2nd Refrigerator | 0.043629 | 0.043629 | 0.043629 | 0.001077 | 0.043629 | 0.043629 | 0.043629 | 0.043629 |
| Coffee Maker | 0.000128 | 1.73E-05 | 0 | 0 | 0 | 5.87E-05 | 0.000594 | 0.000881 |
| Range Oven | 8.46E-05 | 1.14E-05 | 0 | 0 | 0 | 3.89E-05 | 0.000393 | 0.000583 |
| Clothes-washer | 0.000313 | 0 | 0 | 0 | 0 | 0 | 0 | 0.000438 |
| TV | 0.005525 | 0.003153 | 0.001414 | 0.000642 | 0.001414 | 0.001576 | 0.001576 | 0.002373 |
| 2nd TV | 0.000309 | 0.000177 | 7.92E-05 | 0.00025 | 7.92E-05 | 8.83E-05 | 8.83E-05 | 0.000133 |
| Play station | 0.000281 | 0.00016 | 7.18E-05 | 0.000212 | 7.18E-05 | 8E-05 | 8E-05 | 0.00012 |
| Computer | 0.010166 | 0.005801 | 0.002601 | 0.00059 | 0.002601 | 0.0029 | 0.0029 | 0.004365 |
| Air Conditioning | 0.004209 | 0.002195 | 0.00203 | 0.000953 | 0.002195 | 0.002856 | 0.003516 | 0.00586 |
| Hair dryer | 0.0051 | 0.00291 | 0.001305 | 0.000642 | 0.001305 | 0.013455 | 0.016455 | 0.01419 |
| Other Occasional Loads | 0.00153 | 0.000798 | 0.000738 | 0.001025 | 0.000798 | 0.001038 | 0.001278 | 0.00213 |
| Lighting | 0.03825 | 0.01995 | 0.01845 | 0.00615 | 0.01995 | 0.02295 | 0.03195 | 0.06075 |
| Lighting | 0.03825 | 0.01995 | 0.01845 | 0.00615 | 0.01995 | 0.02295 | 0.03195 | 0.06075 |
| Lighting | 0.03825 | 0.01995 | 0.01845 | 0.00615 | 0.01995 | 0.02295 | 0.03195 | 0.06075 |
| Lighting | 0.03825 | 0.01995 | 0.01845 | 0.00615 | 0.01995 | 0.02295 | 0.03195 | 0.06075 |
| Lighting | 0.03825 | 0.01995 | 0.01845 | 0.00615 | 0.01995 | 0.02295 | 0.03195 | 0.06075 |
| Lighting | 0.03825 | 0.01995 | 0.01845 | 0.00615 | 0.01995 | 0.02295 | 0.03195 | 0.06075 |
| Lighting | 0.03825 | 0.01995 | 0.01845 | 0.00615 | 0.01995 | 0.02295 | 0.03195 | 0.06075 |

| Appliance | P start | | | | | | | |
|---|---|---|---|---|---|---|---|---|
| | 9 | 10 | 11 | 12 | 13 | 14 | 15 | 16 |
| Microwave oven | 0.025875 | 0.037125 | 0.043563 | 0.047875 | 0.0495 | 0.044688 | 0.039938 | 0.036813 |
| Refrigerator | 0.140738 | 0.140738 | 0.140738 | 0.140738 | 0.140738 | 0.140738 | 0.140738 | 0.140738 |
| 2nd Refrigerator | 0.043629 | 0.043629 | 0.043629 | 0.043629 | 0.043629 | 0.043629 | 0.043629 | 0.043629 |
| Coffee Maker | 0.00143 | 0.002051 | 0.002407 | 0.002645 | 0.002735 | 0.002469 | 0.002207 | 0.002034 |
| Range Oven | 0.000947 | 0.001358 | 0.001594 | 0.001752 | 0.001811 | 0.001635 | 0.001461 | 0.001347 |
| Clothes-washer | 0.00125 | 0.002881 | 0.004388 | 0.004519 | 0.004519 | 0.004588 | 0.004588 | 0.004588 |
| TV | 0.003949 | 0.005525 | 0.006305 | 0.007881 | 0.007881 | 0.009636 | 0.009961 | 0.01105 |
| 2nd TV | 0.000221 | 0.000309 | 0.000353 | 0.000441 | 0.000441 | 0.00054 | 0.000558 | 0.000619 |
| Play station | 0.0002 | 0.000281 | 0.00032 | 0.0004 | 0.0004 | 0.000489 | 0.000506 | 0.000561 |
| Computer | 0.007266 | 0.010166 | 0.011601 | 0.014502 | 0.014502 | 0.017731 | 0.018329 | 0.020332 |
| Air Conditioning | 0.006719 | 0.006586 | 0.006223 | 0.006553 | 0.006719 | 0.007379 | 0.008204 | 0.009954 |
| Hair dryer | 0.003645 | 0.0051 | 0.00282 | 0.002775 | 0.004275 | 0.002895 | 0.001695 | 0.0027 |
| Other Occasional Loads | 0.002442 | 0.002394 | 0.002262 | 0.002382 | 0.002442 | 0.002682 | 0.002982 | 0.003618 |
| Lighting | 0.04605 | 0.05985 | 0.04905 | 0.0423 | 0.03855 | 0.04905 | 0.05955 | 0.0525 |
| Lighting | 0.04605 | 0.05985 | 0.04905 | 0.0423 | 0.03855 | 0.04905 | 0.05955 | 0.0525 |
| Lighting | 0.04605 | 0.05985 | 0.04905 | 0.0423 | 0.03855 | 0.04905 | 0.05955 | 0.0525 |
| Lighting | 0.04605 | 0.05985 | 0.04905 | 0.0423 | 0.03855 | 0.04905 | 0.05955 | 0.0525 |
| Lighting | 0.04605 | 0.05985 | 0.04905 | 0.0423 | 0.03855 | 0.04905 | 0.05955 | 0.0525 |
| Lighting | 0.04605 | 0.05985 | 0.04905 | 0.0423 | 0.03855 | 0.04905 | 0.05955 | 0.0525 |
| Lighting | 0.04605 | 0.05985 | 0.04905 | 0.0423 | 0.03855 | 0.04905 | 0.05955 | 0.0525 |

| Appliance | P start | | | | | | | |
|---|---|---|---|---|---|---|---|---|
| | 17 | 18 | 19 | 20 | 21 | 22 | 23 | 24 |
| Microwave oven | 0.042375 | 0.046313 | 0.04575 | 0.045188 | 0.043313 | 0.025563 | 0.014375 | 0.006375 |
| Refrigerator | 0.140738 | 0.140738 | 0.140738 | 0.140738 | 0.140738 | 0.140738 | 0.140738 | 0.140738 |
| 2nd Refrigerator | 0.043629 | 0.043629 | 0.043629 | 0.043629 | 0.043629 | 0.043629 | 0.043629 | 0.043629 |
| Coffee Maker | 0.002341 | 0.002559 | 0.002528 | 0.002497 | 0.002393 | 0.001412 | 0.000794 | 0.000352 |
| Range Oven | 0.00155 | 0.001694 | 0.001674 | 0.001653 | 0.001585 | 0.000935 | 0.000526 | 0.000233 |
| Clothes-washer | 0.004644 | 0.004644 | 0.004838 | 0.004838 | 0.004644 | 0.003825 | 0.002438 | 0.000563 |
| TV | 0.01105 | 0.01105 | 0.012626 | 0.013406 | 0.01105 | 0.008678 | 0.007881 | 0.006289 |
| 2nd TV | 0.000619 | 0.000619 | 0.000707 | 0.000751 | 0.000619 | 0.000486 | 0.000441 | 0.000352 |
| Play station | 0.000561 | 0.000561 | 0.000641 | 0.000681 | 0.000561 | 0.000441 | 0.0004 | 0.000319 |
| Computer | 0.020332 | 0.020332 | 0.023232 | 0.024668 | 0.020332 | 0.015967 | 0.014502 | 0.011571 |
| Air Conditioning | 0.010433 | 0.011291 | 0.012117 | 0.01248 | 0.011209 | 0.011011 | 0.00799 | 0.005315 |
| Hair dryer | 0.0057 | 0.0072 | 0.011655 | 0.012375 | 0.0102 | 0.00801 | 0.007275 | 0.005805 |
| Other Occasional Loads | 0.003792 | 0.004104 | 0.004404 | 0.004536 | 0.004074 | 0.004002 | 0.002904 | 0.001932 |
| Lighting | 0.0903 | 0.11535 | 0.1251 | 0.1284 | 0.1296 | 0.12255 | 0.09735 | 0.06375 |
| Lighting | 0.0903 | 0.11535 | 0.1251 | 0.1284 | 0.1296 | 0.12255 | 0.09735 | 0.06375 |
| Lighting | 0.0903 | 0.11535 | 0.1251 | 0.1284 | 0.1296 | 0.12255 | 0.09735 | 0.06375 |
| Lighting | 0.0903 | 0.11535 | 0.1251 | 0.1284 | 0.1296 | 0.12255 | 0.09735 | 0.06375 |
| Lighting | 0.0903 | 0.11535 | 0.1251 | 0.1284 | 0.1296 | 0.12255 | 0.09735 | 0.06375 |
| Lighting | 0.0903 | 0.11535 | 0.1251 | 0.1284 | 0.1296 | 0.12255 | 0.09735 | 0.06375 |
| Lighting | 0.0903 | 0.11535 | 0.1251 | 0.1284 | 0.1296 | 0.12255 | 0.09735 | 0.06375 |

### E. Load Profiling of Four Rooms Public Housing

For brevity purpose, for the four rooms flats, the starting probablity is shown in Table VI, and the daily load profile is shown in Fig. 8, giving an average value of 347.7 kWh. From EMA report [1], the average monthly electricity consumption in 4 rooms flat is 379.6 kWh (see Section II-A).

## V. VALIDATION WITH MEASURED DATA

### A. Real-time Energy Portal at NTU, Singapore

The Energy Market Authority (EMA), together with Singapore Power (SP), carried out the Intelligent Energy System (IES) pilot project in 2009 [25]. Fitted with smart meters, electricity usage of 323 staff house units at Nanyang Technological University (NTU), Singapore has been monitored and recorded, using a web-based Customer Energy Portal (CEP), every 30 min since June 2012 until now [25]–[27]. Fig. 9 shows the automated metering infrastructure implemented in the CEP at NTU [26].

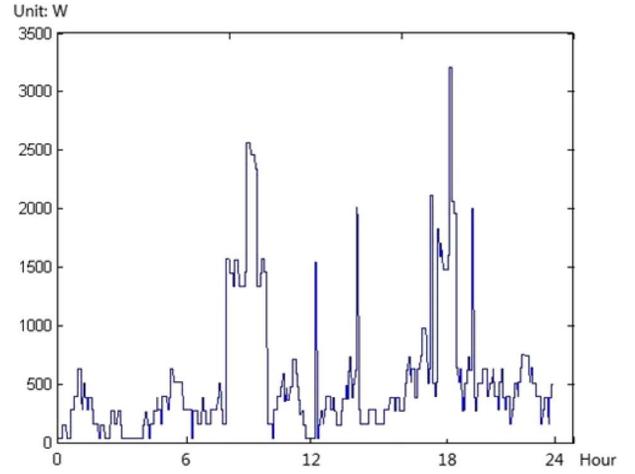

Fig. 6. Daily load profile of five rooms flat.

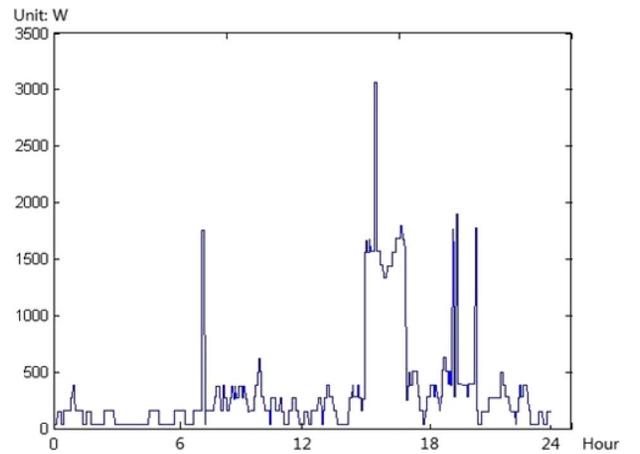

Fig. 7. Daily load profile of three rooms flat.

### B. Validation

From the NTU CEP, a random day in February 2013 was selected and the daily average load is shown in Fig. 10. According to the NTU monitored data, the average load per unit is 220 kWh. According to the EMA statistics (see Section II-A), the average of 1 or 2-rooms (153.8 kWh) and 3-rooms (278.2 kWh) public housing load is 216 kWh. In fact, the majority of the monitored units in NTU are 3-room flats and the rests are 2-room flats. This observation implies that the load profile of households in NTU complies with the average statistics of Singapore.

From Fig. 10, two electricity load peaks can be observed at 07:30-08:30 h, and 20:30-21:30 h. The load is relatively low in the early morning and from 09:30 until 17:30 in a day. This is fairly reasonable as activities involving electricity are more when people are at home in the morning and evening.

Comparing Figs. 7 and 10, we can notice the similarities between the two load profiles, in terms of peak profiles and average energy utilizations. However, there are minor differences,

TABLE VI
Starting Probability for 4 Rooms Flat

| Appliance | P start | | | | | | | |
|---|---|---|---|---|---|---|---|---|
| | 1 | 2 | 3 | 4 | 5 | 6 | 7 | 8 |
| Microwave oven | 0.001721 | 0.000233 | 0 | 0 | 0 | 0.000791 | 0.007998 | 0.011858 |
| Refrigerator | 0.140738 | 0.140738 | 0.140738 | 0.003475 | 0.140738 | 0.140738 | 0.140738 | 0.140738 |
| 2nd Refrigerator | 0.021111 | 0.021111 | 0.021111 | 0.000521 | 0.021111 | 0.021111 | 0.021111 | 0.021111 |
| Coffee Maker | 0.000103 | 1.39E-05 | 0 | 0 | 0 | 4.72E-05 | 0.000478 | 0.000708 |
| Range Oven | 7.25E-05 | 9.8E-06 | 0 | 0 | 0 | 3.33E-05 | 0.000337 | 0.0005 |
| Clothes-washer | 0.000174 | 0 | 0 | 0 | 0 | 0 | 0 | 0.000243 |
| TV | 0.005525 | 0.003153 | 0.001414 | 0.000642 | 0.001414 | 0.001576 | 0.001576 | 0.002373 |
| 2nd TV | 0.000167 | 9.51E-05 | 4.26E-05 | 0.000135 | 4.26E-05 | 4.75E-05 | 4.75E-05 | 7.15E-05 |
| Play station | 0.000221 | 0.000126 | 5.66E-05 | 0.000167 | 5.66E-05 | 6.31E-05 | 6.31E-05 | 9.49E-05 |
| Computer | 0.008341 | 0.004759 | 0.002134 | 0.00059 | 0.002134 | 0.00238 | 0.00238 | 0.003582 |
| Air Conditioning | 0.003086 | 0.001609 | 0.001488 | 0.000902 | 0.001609 | 0.002093 | 0.002577 | 0.004296 |
| Hair dryer | 0.004743 | 0.002706 | 0.001214 | 0.000597 | 0.001214 | 0.012513 | 0.015303 | 0.013197 |
| Other Occasional Loads | 0.001424 | 0.000743 | 0.000687 | 0.001025 | 0.000743 | 0.000966 | 0.001189 | 0.001982 |
| Lighting | 0.03825 | 0.01995 | 0.01845 | 0.00615 | 0.01995 | 0.02295 | 0.03195 | 0.06075 |
| Lighting | 0.03825 | 0.01995 | 0.01845 | 0.00615 | 0.01995 | 0.02295 | 0.03195 | 0.06075 |
| Lighting | 0.03825 | 0.01995 | 0.01845 | 0.00615 | 0.01995 | 0.02295 | 0.03195 | 0.06075 |
| Lighting | 0.03825 | 0.01995 | 0.01845 | 0.00615 | 0.01995 | 0.02295 | 0.03195 | 0.06075 |
| Lighting | 0.03825 | 0.01995 | 0.01845 | 0.00615 | 0.01995 | 0.02295 | 0.03195 | 0.06075 |
| Lighting | 0.03825 | 0.01995 | 0.01845 | 0.00615 | 0.01995 | 0.02295 | 0.03195 | 0.06075 |

| Appliance | P start | | | | | | | |
|---|---|---|---|---|---|---|---|---|
| | 9 | 10 | 11 | 12 | 13 | 14 | 15 | 16 |
| Microwave oven | 0.019251 | 0.027621 | 0.032411 | 0.035619 | 0.036828 | 0.033248 | 0.029714 | 0.027389 |
| Refrigerator | 0.140738 | 0.140738 | 0.140738 | 0.140738 | 0.140738 | 0.140738 | 0.140738 | 0.140738 |
| 2nd Refrigerator | 0.021111 | 0.021111 | 0.021111 | 0.021111 | 0.021111 | 0.021111 | 0.021111 | 0.021111 |
| Coffee Maker | 0.00115 | 0.001649 | 0.001935 | 0.002127 | 0.002199 | 0.001985 | 0.001774 | 0.001635 |
| Range Oven | 0.000811 | 0.001164 | 0.001366 | 0.001501 | 0.001552 | 0.001401 | 0.001252 | 0.001154 |
| Clothes-washer | 0.000695 | 0.001602 | 0.00244 | 0.002513 | 0.002513 | 0.002551 | 0.002551 | 0.002551 |
| TV | 0.003949 | 0.005525 | 0.006305 | 0.007881 | 0.007881 | 0.009636 | 0.009961 | 0.01105 |
| 2nd TV | 0.000119 | 0.000167 | 0.00019 | 0.000238 | 0.000238 | 0.000291 | 0.0003 | 0.000333 |
| Play station | 0.000158 | 0.000221 | 0.000252 | 0.000315 | 0.000315 | 0.000385 | 0.000398 | 0.000442 |
| Computer | 0.005962 | 0.008341 | 0.009519 | 0.011899 | 0.011899 | 0.014548 | 0.015039 | 0.016683 |
| Air Conditioning | 0.004925 | 0.004828 | 0.004562 | 0.004804 | 0.004925 | 0.005409 | 0.006014 | 0.007296 |
| Hair dryer | 0.00339 | 0.004743 | 0.002623 | 0.002581 | 0.003976 | 0.002692 | 0.001576 | 0.002511 |
| Other Occasional Loads | 0.002272 | 0.002228 | 0.002105 | 0.002217 | 0.002272 | 0.002496 | 0.002775 | 0.003367 |
| Lighting | 0.04605 | 0.05985 | 0.04905 | 0.0423 | 0.03855 | 0.04905 | 0.05955 | 0.0525 |
| Lighting | 0.04605 | 0.05985 | 0.04905 | 0.0423 | 0.03855 | 0.04905 | 0.05955 | 0.0525 |
| Lighting | 0.04605 | 0.05985 | 0.04905 | 0.0423 | 0.03855 | 0.04905 | 0.05955 | 0.0525 |
| Lighting | 0.04605 | 0.05985 | 0.04905 | 0.0423 | 0.03855 | 0.04905 | 0.05955 | 0.0525 |
| Lighting | 0.04605 | 0.05985 | 0.04905 | 0.0423 | 0.03855 | 0.04905 | 0.05955 | 0.0525 |
| Lighting | 0.04605 | 0.05985 | 0.04905 | 0.0423 | 0.03855 | 0.04905 | 0.05955 | 0.0525 |

| Appliance | P start | | | | | | | |
|---|---|---|---|---|---|---|---|---|
| | 17 | 18 | 19 | 20 | 21 | 22 | 23 | 24 |
| Microwave oven | 0.031527 | 0.034457 | 0.034038 | 0.03362 | 0.032225 | 0.019019 | 0.010695 | 0.004743 |
| Refrigerator | 0.140738 | 0.140738 | 0.140738 | 0.140738 | 0.140738 | 0.140738 | 0.140738 | 0.140738 |
| 2nd Refrigerator | 0.021111 | 0.021111 | 0.021111 | 0.021111 | 0.021111 | 0.021111 | 0.021111 | 0.021111 |
| Coffee Maker | 0.001883 | 0.002058 | 0.002033 | 0.002008 | 0.001924 | 0.001136 | 0.000639 | 0.000283 |
| Range Oven | 0.001329 | 0.001452 | 0.001435 | 0.001417 | 0.001358 | 0.000802 | 0.000451 | 0.0002 |
| Clothes-washer | 0.002583 | 0.002583 | 0.00269 | 0.00269 | 0.002583 | 0.002127 | 0.001356 | 0.000313 |
| TV | 0.01105 | 0.01105 | 0.012626 | 0.013406 | 0.01105 | 0.008678 | 0.007881 | 0.006289 |
| 2nd TV | 0.000333 | 0.000333 | 0.000381 | 0.000404 | 0.000333 | 0.000262 | 0.000238 | 0.00019 |
| Play station | 0.000442 | 0.000442 | 0.000505 | 0.000536 | 0.000442 | 0.000347 | 0.000315 | 0.000252 |
| Computer | 0.016683 | 0.016683 | 0.019062 | 0.02024 | 0.016683 | 0.013101 | 0.011899 | 0.009494 |
| Air Conditioning | 0.007647 | 0.008276 | 0.008881 | 0.009148 | 0.008216 | 0.008071 | 0.005856 | 0.003896 |
| Hair dryer | 0.005301 | 0.006696 | 0.010839 | 0.011509 | 0.009486 | 0.007449 | 0.006766 | 0.005399 |
| Other Occasional Loads | 0.003529 | 0.003819 | 0.004098 | 0.004221 | 0.003791 | 0.003724 | 0.002702 | 0.001798 |
| Lighting | 0.0903 | 0.11535 | 0.1251 | 0.1284 | 0.1296 | 0.12255 | 0.09735 | 0.06375 |
| Lighting | 0.0903 | 0.11535 | 0.1251 | 0.1284 | 0.1296 | 0.12255 | 0.09735 | 0.06375 |
| Lighting | 0.0903 | 0.11535 | 0.1251 | 0.1284 | 0.1296 | 0.12255 | 0.09735 | 0.06375 |
| Lighting | 0.0903 | 0.11535 | 0.1251 | 0.1284 | 0.1296 | 0.12255 | 0.09735 | 0.06375 |
| Lighting | 0.0903 | 0.11535 | 0.1251 | 0.1284 | 0.1296 | 0.12255 | 0.09735 | 0.06375 |
| Lighting | 0.0903 | 0.11535 | 0.1251 | 0.1284 | 0.1296 | 0.12255 | 0.09735 | 0.06375 |

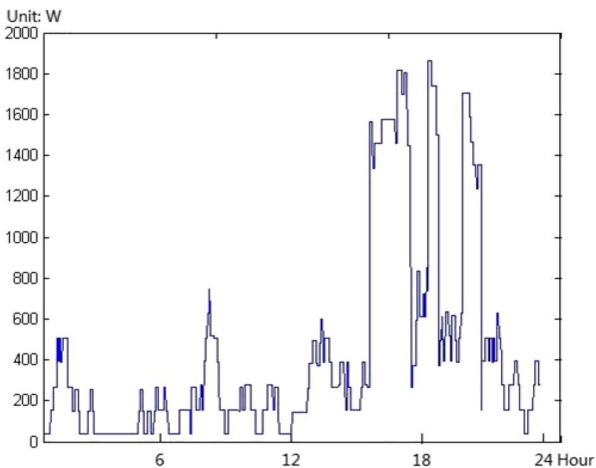

Fig. 8. Daily load profile of four rooms flat.

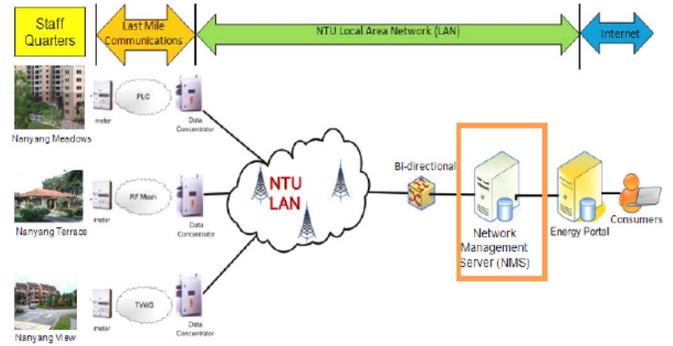

Fig. 9. Automated metering infrastructure in the energy portal at NTU.

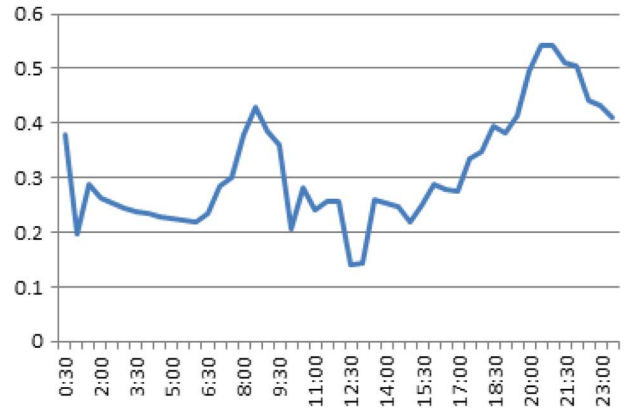

Fig. 10. Average residential load in NTU, Singapore on February 6, 2013.

as Fig. 7 is modeled for public housing, while Fig. 10 is from NTU, a university. Being a university, the building load patterns vary slightly. However, overall quite good similarity could be observed for the modeled profile and the measured profile. This substantiates the bottom-up load profiling approach.

## VI. Discussions

The following comments are cited on the results.
1) Four types of households were considered and all simulated loads tally with the statistical data provided by EMA, Singapore [1].
2) The simulated load profiled match quite well with the measured value from NTU campus houses.
3) However, it is to be noted that for validation using real-measurements, campus housing in NTU is used, most of which is 3 room-units, quite similar to public housing. This is because detailed energy account for all public houses in Singapore is usually unavailable. On the other hand, the real-time detailed energy utilization statistics is available from NTU CEP. Therefore, the validation using NTU CEP data should be considered as corresponding to the 3-room public housing units for Singapore.
4) One of the main differences between the NTU CEP data (e.g., Fig. 10) with the public housing (e.g., Fig. 7) is that in the NTU CEP data, there is a peak around 9:30 PM. This is mainly because of typical university activity, e.g., evening studies, etc. by the students in the campus housing units.

Compared to this, there is afternoon peaks around 5 PM in the public housing (Fig. 7), as people normally return from work.

5) Variations in the load profile patterns for houses with different number of rooms can be observed. For example, in Fig. 6 for 5 rooms houses, some peaks could be seen between 12:00-14:00 h. These are absent in Figs. 5 and 7 for 1 or 2 rooms and 3 rooms flats. This is because for a 5 room house, probability of electrical appliance usage is higher due to expectedly higher number of inhabitants.
6) In other load profiling studies in other countries, typically seasonal variation could be observed. However, as the daily temperature pattern and daylight hours change slightly in Singapore, the daily electricity consumption on a yearly level can be considered as consistent with insignificant change.
7) Effect of ambient temperatures on energy consumption can be seen, but not very dominant for Singapore due to its tropical climate.
8) Besides weather condition, other external variables that influence the demand include income, demographic characteristics, etc. These are factored in the different unit-size ranging from 1 room to 5 rooms flats considered in this study. Annual report by the Singapore Housing Development Board (HDB) [2] supports the fact that the public housing size is a good indicator of demographic and economical conditions in Singapore.
9) Uncertain factors such as abnormal weather, national or international events have not been considered in this study. Also, differentiation between weekdays and weekends are not considered. These would be taken into account in future studies.
10) A bottom-up mathematical model is established to simulate the residential building load profile. Bottom-up approach uses statistical data for simulating households' electricity consumption. In this approach, the total load in the building is built up from the elementary load components, which can be a single household or even a single piece of appliance. The advantage of bottom-up approach is that it can analyze every individual appliance's effect on the total load, which could be quite helpful in the future study of smart grid.

## VII. Conclusion

The goal of this study is to classify different types of residential buildings in Singapore and to study the load profile of each group. Residential building is classified into 1 or 2-room unit, 3-room unit, 4-room unit and 5-room unit and the power load profile for each type of residential building is simulated and verified by comparing with statistical data on average monthly usage published by Energy Market Authority. A bottom-up mathematical model is established to simulate the residential building load profile. The detailed mathematical model considers typical electrical appliances, its relative saturation depending on the building type, the power rating and the utilization pattern in terms of frequency of operation, probability of being on at particular time, etc. MATLAB is used to simulate the models and create the different load profiles. The simulated models are successfully verified against measured data, using a web-based Customer Energy Portal at the campus housings of Nanyang Technological University, Singapore. The detailed modeling of the electricity consumption habit for each type of household could be helpful in future research, especially for energy efficient operations in buildings, which is a key area in the smart grid initiative.

ACKNOWLEDGMENT

The authors would like to thank Mr. C. P. Huat for providing the assistance in retrieving NTU monitored energy consumption data.

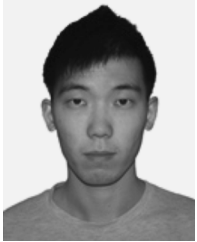

**Luo Chuan** received the B.Eng. degree in electrical and electronic engineering (EEE) from the Nanyang Technological University (NTU), Singapore, in 2013. He is pursuing the M.Sc. degree (power engineering) in the School of EEE at NTU.

Currently, he is a full-time engineer at SingaporePower. His research interest includes power system, energy modeling, and energy efficiency.

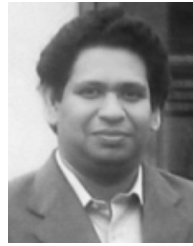

**Abhisek Ukil** (S'05–M'06–SM'10) received the B.E. degree in electrical engineering from the Jadavpur University, Kolkata, India, in 2000 and the M.Sc. degree in electronic systems and engineering management from the University of Bolton, Bolton, U.K., in 2004. He received the Ph.D. degree from the Pretoria (Tshwane) University of Technology, Pretoria, South Africa, in 2006, working on automated disturbance analysis in power systems.

Since 2013, he has been an Assistant Professor in the School of Electrical and Electronic Engineering, Nanyang Technological University, Singapore. From 2006–2013, he was Principal Scientist at the ABB Corporate Research Center, Baden-Daettwil, Switzerland. He is author/coauthor of more than 60 refereed papers, a monograph, 2 chapters, and inventor/co-inventor of 9 patents. His research interests include smart grid, power systems, HVDC, renewable energy & integration, condition monitoring, and signal processing applications.